\documentclass[aps,pre,twocolumn,amsmath,amsfonts,amssymb,floatfix,superscriptaddress,nofootinbib,notitlepage,10pt]{revtex4-1}
\usepackage{amssymb}
\usepackage{amsmath}
\usepackage{graphicx,color}
\usepackage{bbm}
\usepackage{multirow}
\usepackage[colorlinks,linkcolor=blue,urlcolor=blue]{hyperref}

\allowdisplaybreaks

\newcommand{\beq}{\begin{equation}}
\newcommand{\eeq}{\end{equation}}
\newcommand{\beqnn}{\begin{equation*}}
\newcommand{\eeqnn}{\end{equation*}}
\newcommand{\bea}{\begin{eqnarray}}
\newcommand{\eea}{\end{eqnarray}}
\newcommand{\beann}{\begin{eqnarray*}}
\newcommand{\eeann}{\end{eqnarray*}}

\begin{document}

\title{Particle pinning suppresses spinodal criticality in the shear banding instability}

\author{Bhanu Prasad Bhowmik}
\affiliation{Tata Institute of Fundamental Research, 36/P, Gopanpally Village, Serilingampally Mandal,Ranga Reddy District, Hyderabad, 500107, Telangana, India}

\author{Smarajit Karmakar}
\affiliation{Tata Institute of Fundamental Research, 36/P, Gopanpally Village, Serilingampally Mandal,Ranga Reddy District, Hyderabad, 500107, Telangana, India}

\author{Itamar Procaccia}
\affiliation{Department of chemical and biological physics, the Weizmann Institute of Science, Rehovot 76100, Israel}

\author{Corrado Rainone}
\affiliation{Institute for Theoretical Physics, University of Amsterdam, Science Park 904, 1098 XH Amsterdam, The Netherlands}

\begin{abstract}
Strained amorphous solids often fail mechanically by creating a shear-band. It had been understood that the shear banding instability is usefully described as crossing a spinodal point (with disorder) in an appropriate thermodynamic description. It remained contested however whether the spinodal is critical (with divergent correlation length) or not. Here we offer evidence for
critical spinodal by using particle pinning. For a finite concentration of pinned particles the correlation
length is bounded by the average distance between pinned particles, but without pinning it is bounded by the system size.
\end{abstract}

\maketitle

\section{Introduction}
When strained, perfect crystalline materials exhibit an elastic stress response all the way to mechanical yield~\cite{AshcroftMermin}, whereupon the material start to flow plastically. In contradistinction, in the case of amorphous solids the response to strain is already plastic at small strains.  In the thermodynamic limit plasticity exists at arbitrarily small strains ~\cite{KLP10b,11HKLP,LPRS18}) . The stress vs. strain curve is typically punctuated by stress drops due to plastic instabilities~\cite{04VBB,04ML,05DA,06TLB,06ML,09LP,11RTV}. In this case, elasticity cannot be clearly decoupled from plasticity~\cite{11HKLP} and the definition of a mechanical yield transition becomes a subtle problem.

Help comes upon considering the non-affine particle rearrangements that accompany plastic instabilities, and in particular their system-size scaling. For model amorphous solids under Athermal Quasi-Static (AQS) strain it is now known that the energy drops in the pre-yield phase (wherein the stress response to strain is still growing on the average) do not scale with the system size~\cite{KLP10b}. This is consistent with the non-affine particle rearrangements being localized in space. After yield, in the plastic steady state which is found at high strain (wherein the response is on average constant), on the other hand, the energy drops exhibit a scaling $\Delta U \propto N^{1/3}$, irrespectively of the model or the dimensionality of space~\cite{KLP10b}. The corresponding particle rearrangements become extended, assuming the form of micro shear-bands~\cite{12DHP} which induce a global collapse of the sample. Such a collapse has huge relevance for the science of brittle materials, such as metallic glasses~\cite{06AG,SW15,Johnson99}, which, despite sometimes exhibiting a toughness higher than ordinary crystalline metallic alloys, cannot be normally employed as structural materials because of their tendency to fail catastrophically through shear banding~\cite{12DHP,13DHP,13DGMPS}.

The upshot is that amorphous yield can therefore be visualized as a transition from localized to extended plastic activity. As for the nature of this transition, however, an established consensus is still missing. One point of view, advanced by some of us, advocates a transition in the form of a critical spinodal with disorder~\cite{16JPRS,17PPRS,17PRS}, wherein the growth in space of plastic events is controlled by a correlation length which diverges at the transition and can be measured from suitable multi-point correlators~\cite{17PPRS,17PRS}; another one~\cite{OBBRT18,PGW18}, inspired by previous work~\cite{16NBT} on disordered spinodals in the context of the Random Field Ising Model~\cite{N98,PDS99,SDKKRS93} supports a picture in terms, again, of a spinodal transition, however not a critical one. In this view avalanches grow through nucleation around rare defects, rather than through a critical process involving a correlation length. Besides the obvious difference between these two interpretations in terms of physics, there is also a genuinely practical one, namely the possibility (or not) of identifying precursors to mechanical yield, which is present in the first picture but not in the second one. As of now, it still remains to be seen which scenario provides the best description of the physics of yielding, a verification which requires a systematic testing of these ideas across different material preparation protocols and model systems.

In a recent work~\cite{13DMPS,PCK18}, it was shown how this phenomenology is qualitatively changed, once inclusions in the form of pinned particles~\cite{CB12,KB13,OKIM15,CKD15-1,CKD15-2,DCK17} are added to the glass. While the glass still yields, the growth of plastic activity is suppressed to the point that events remain localized, even in the plastic steady-state. Moreover, the shear modulus increases and a prominent stress overshoot appears. Therefore, in this case, mechanical yield ceases to be a sharp transition mediated by the appearance of material-collapsing shear bands, but rather takes place through progressive accumulation of \emph{localized} plastic activity.

In this work, we apply the techniques formulated in~\cite{16JPRS,17PPRS,17PRS} to the study of mechanical yield in such a model system, and use them to elucidate the mechanism leading to the suppression of shear bands: while a growing correlation length can still be measured, we show that it cannot diverge in a critical fashion, and is on the contrary bounded by the typical distance between pinned particles. Our picture of yield in terms of a critical spinodal is therefore able to capture and explain the qualitative changes in the yield and mechanics of glasses with pinned particles. As a byproduct, this work and refs.~\cite{PCK18,HMPS16} introduce and justify theoretically a protocol which can, in principle, be adapted to stymie shear banding and brittleness in material science applications, including metallic glasses.

%, whilst in the second case it is not the case. It goes without saying that the possibility of finding tell-tale precursors to yield is relevant from problems that go from the prediction of earthquakes in geology, to the mechanics of brittle materials, such %as metallic glasses~\cite{06AG,SW15,Johnson99}, which, despite sometimes  exhibiting a toughness comparable to the one of ordinary, crystalline metallic alloys, cannot be normally employed as structural materials because of their tendency to %fail catastrophically and without any evident precursors~\cite{12DHP,13DHP,13DGMPS}.

\section{Materials and methods}

\paragraph{Model}
We have performed simulations of a canonical glass forming system, the Kob-Andersen model in two dimensions~\cite{94KA}. It is a binary mixture with 65\% large particles (type A) and 35\% small particles (type B). The particles interact via the Lennard-Jones (LJ) potential
\begin{equation}
V_{\alpha,\beta} = 4\epsilon_{\alpha\beta} [(\sigma_{\alpha\beta}/r)^{12} - (\sigma_{\alpha\beta}/r)^6],
\end{equation}
where $\alpha$ and $\beta$ stand for the two different types of particle A and B.
$\epsilon_{AA}$ = 1.0, $\epsilon_{AB}$ = 1.50, $\epsilon_{BB}$ = 0.50.
$\sigma_{AA}$ = 1.0, $\sigma_{AB}$ = 0.88, $\sigma_{BB}$ = 0.80.

\paragraph{Sample preparation}
To prepare the initial state, the system is first equilibrated at a temperature $T = 0.4$,  then  brought down to a lower temperature $T = 10^{-5}$ using a cooling rate $\dot{T} = 10^{-5}$. After that a fraction of particles are chosen at random and pinned in place. Once pinned, these particles do not participate further in the dynamics and stay frozen in their respective positions. Afterwards, the system is heated up to a temperature $T = 0.2$, and  then, in order to produce the desired number of replicated configurations, a snapshot of the positions is taken and a set of particle velocities, drawn from the Maxwell-Boltzmann distribution at $T=0.2$, is assigned to it. We repeat this procedure as many times as the number of replicas we wish to employ to measure the correlators. All the so obtained initial configurations are then cooled to $T = 0$ with a cooling rate $10^{-1}$, in order to finally obtain our athermal samples. We restate that all of these replicas share the same configuration of the pinned particles. We perform this procedure on systems of $N=500,\ 1000,\ 2000,\ 4000\ \textrm{and}\ 10000$ particles. We use $35$ patches (each with $150$ replicas), $30$ patches (each with $120$ replicas), $20$ patches (each with $30$ replicas), $20$ patches (each with $60$ replicas) and $20$ patches(each with $48$ replicas) for $N=500,\ 1000,\ 2000,\ 4000\ \textrm{and}\ 10000$ respectively. 

\paragraph{Shear Protocol}
We then perform an athermal quasi-static (AQS) shear protocol on these inherent states. As usual in AQS, we strain the system by first applying an affine shear transformation $\delta\gamma = 5\times10^{-5}$, and then the potential energy is minimized using Conjugate Gradient in order to return the configuration to mechanical equilibrium. During the shearing process, the pinned particles undergo an affine transformation like the unpinned ones, but they remain frozen in place during the minimization step.

\paragraph{Pinning Protocol}
As the particles do not move during minimization, it is possible that the affine step bring them close to each other, with the result of making the potential energy unphysically large. To avoid this situation, the randomly pinned particles are chosen in such a way that the following condition is satisfied: $y_i-y_j \ge r_c$, where $y_i$ is the y component of the position vector of $i^{th}$ particle. $r_c$ is the distance where the repulsive part of the interaction vanishes.
\begin{figure}
 \includegraphics[width=0.30\textwidth]{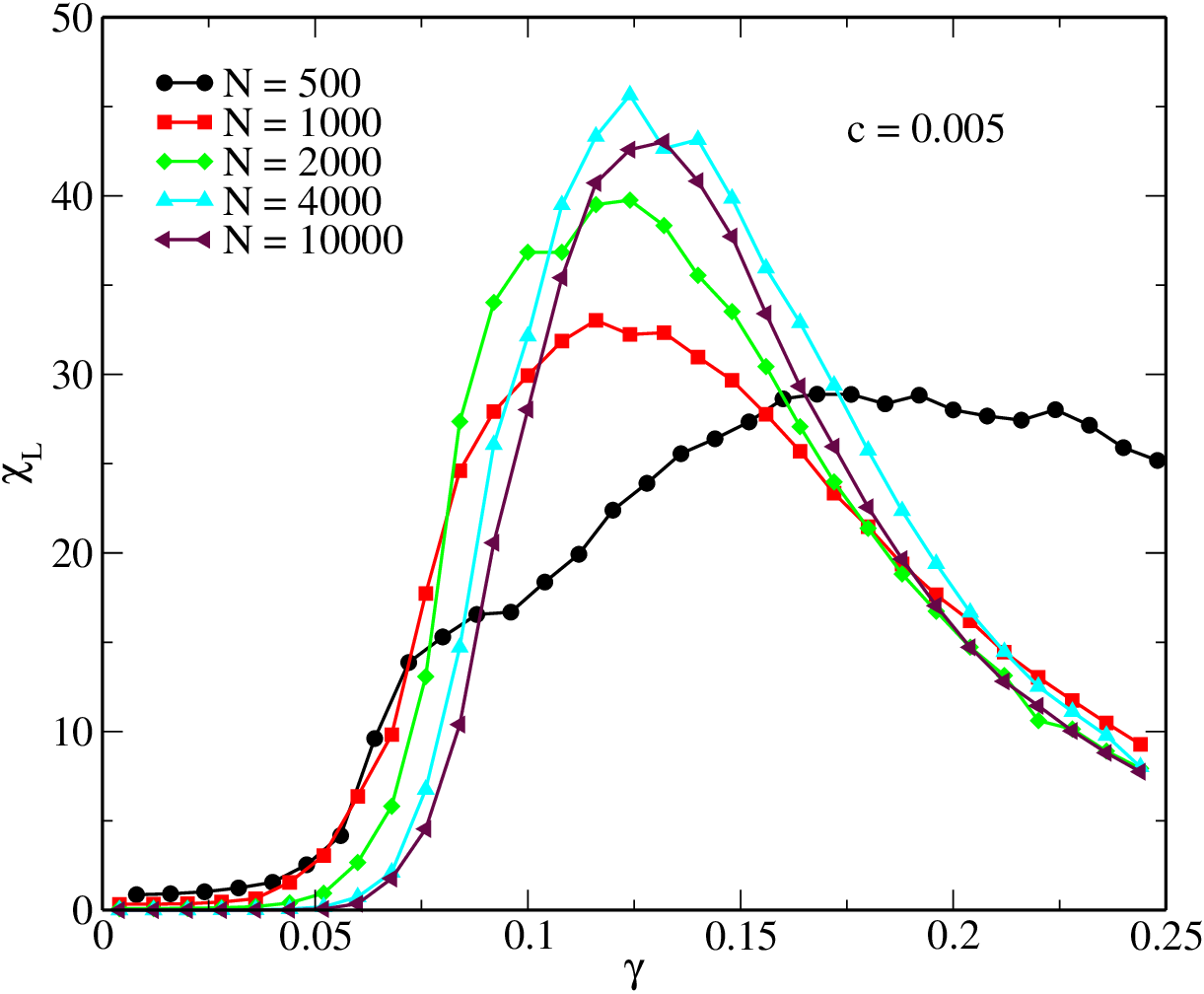}
 \includegraphics[width=0.30\textwidth]{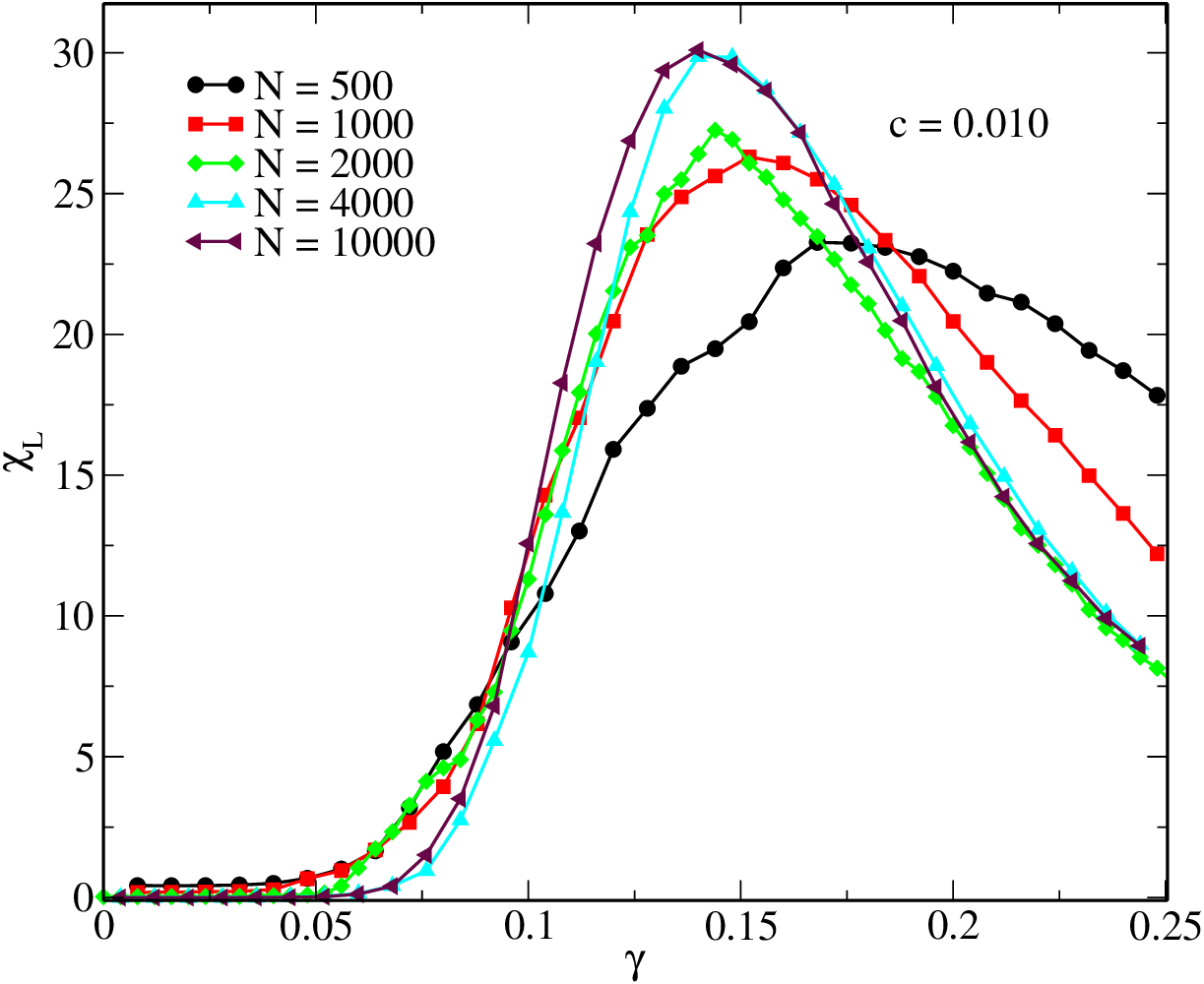}
 \includegraphics[width=0.30\textwidth]{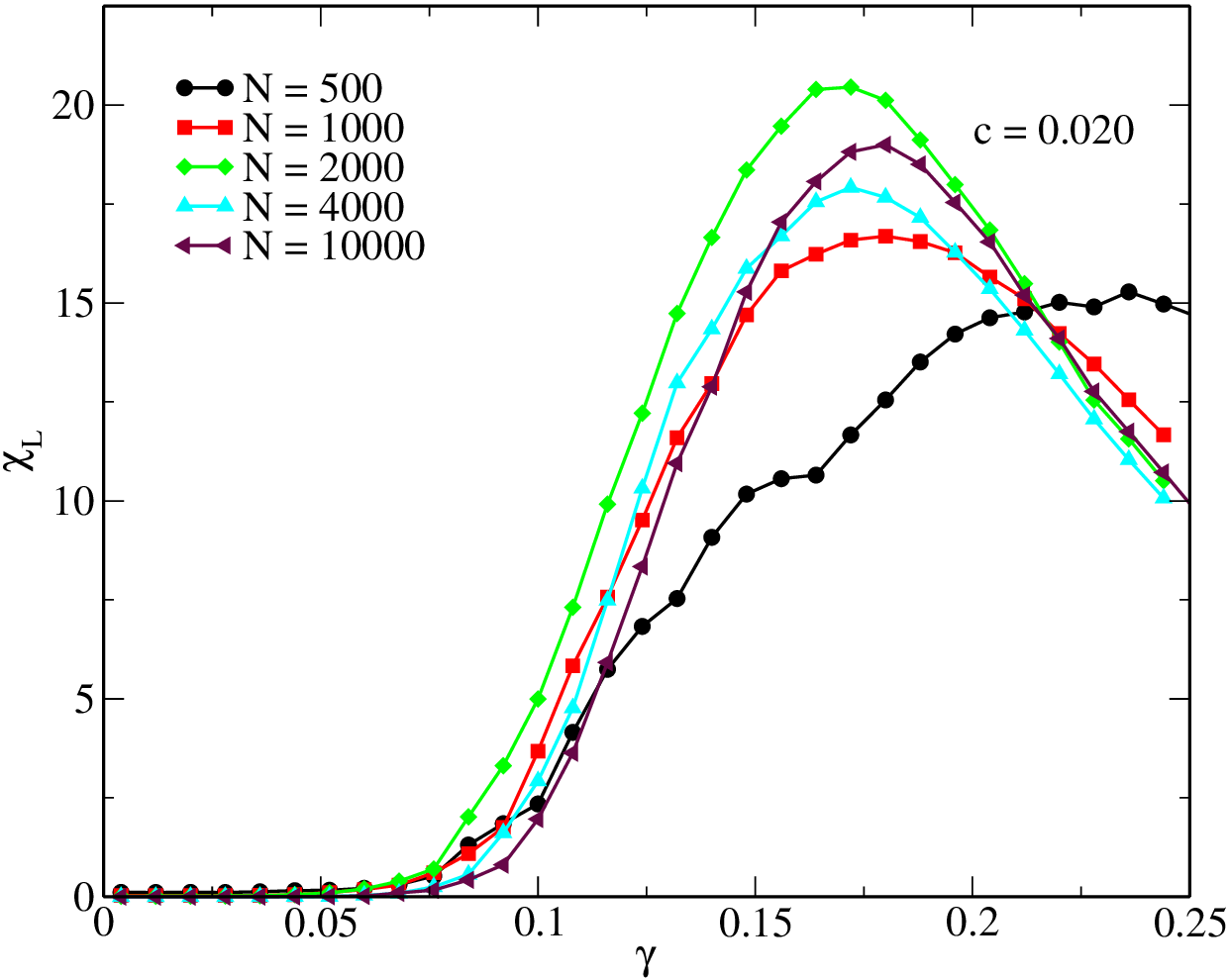}
 \includegraphics[width=0.30\textwidth]{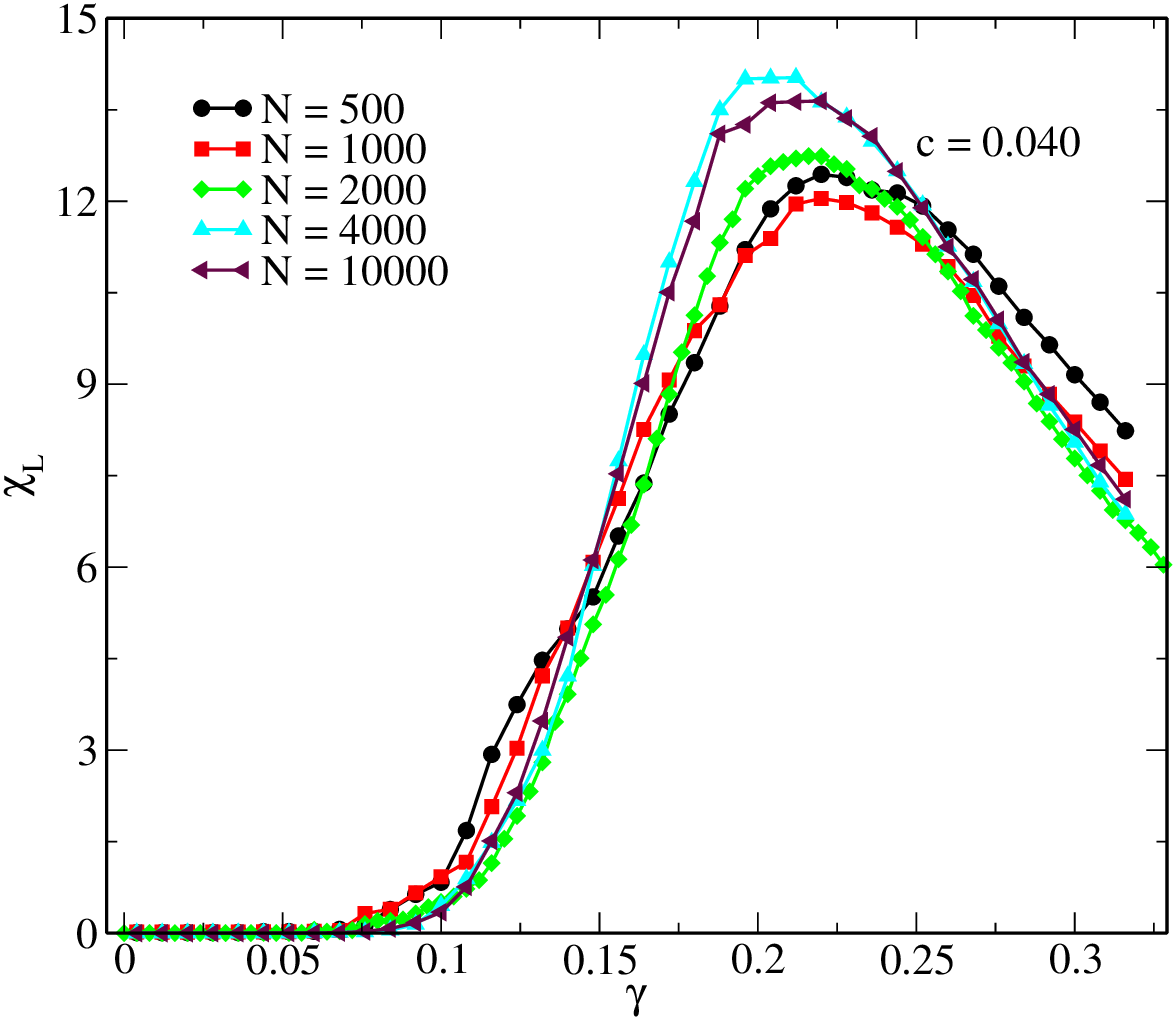}
 \caption{The replica susceptibility $\chi_{_L}$, for four different values of the pinning concentration $c$. Even at the smallest value corresponding to just 0.5\% of pinned particles, the susceptibility peak is not seen diverging as the system size grows. \label{fig:thesusc}}
\end{figure}

\section{Results}
Once our ensembles of replicas are created, we then calculate the multi-point correlators defined in appendix \ref{app:corrs} and compute the associated susceptibility $\chi_L$ (denoted as $\chi_{G_L}$ in ref.~\cite{17PRS}) as a function of the shear strain. In Fig.~\ref{fig:thesusc} we report the results for all system sizes.

It emerges that the inclusion of pinned particles into the system at \emph{any} finite concentration destroys the critical behavior of the yielding transition. The susceptibility peaks as the system yields, but this peak does not grow with system size as in the unpinned case (see e.g.~\cite[Fig. 10]{17PRS}); a much milder growth, without hints to a divergence, is visible, whose theoretical justification is the subject of the next section. Besides, we also observe that the yielding point itself is shifted to higher strains, consistently with the results of ref.~\cite{13DMPS,PCK18}.

\section{Theoretical Considerations}
We provide now a simple theory of the mild and seemingly non-critical growth of the susceptibility peak. We shall simply assume that the shear-banding correlation length~\cite{17PPRS,17PRS} which controls the scale of avalanches on approaching yield, cannot grow beyond the typical distance between pinned particles. This implies that at the susceptibility peak, corresponding to the maximal value of the correlation length, one must have
\begin{equation}
 \xi^* \simeq \frac{1}{(c\rho)^{1/d}},
\label{eq:xistar}
\end{equation}
with $\rho$ being the number density of the system.

Let us assume that the replica correlator $G_L(r)$ has the standard form for a correlation function in critical phenomena~\cite{02Z-J}
$$
G_L(r) = \frac{e^{-r/\xi}}{r^{d-2+\eta}},
$$
where $d$ is the dimension of space and $\eta$ the anomalous dimension. One then has for the associated susceptibility
$$
\chi_{_L} \propto \int_0^L dr\ r^{1-\eta}e^{-r/\xi}.
$$
Where $L=V^{1/d}$ is the linear size of the system. The integral can be calculated analytically, obtaining
\beq
\chi_{_L} \propto L^{-\eta} \left(\frac{L}{\xi}\right)^{\eta} \xi^2 \left[\Gamma(2-\eta) -\Gamma\left(2-\eta,\frac{L}{\xi}\right)\right],
\label{eq:susc}
\eeq
with the definitions of the Euler Gamma and Incomplete Gamma functions
\beq
\begin{split}
\Gamma(x) \equiv &\int_{0}^\infty dt\ t^{x-1}e^{-t},\\
\Gamma(x,a) \equiv &\int_{a}^{\infty}dt\ t^{x-1}e^{-t}.
\end{split}
\eeq
If one takes the thermodynamic limit, using $\lim_{a\to\infty}\Gamma(x,a) = 0$ one obtains
\beq
\chi_{_L}\propto \xi^{2-\eta}\Gamma(2-\eta),
\eeq
which yields the well-known scaling relation between the susceptibility and lengthscale critical exponents $\gamma$ and $\nu$~\cite{02Z-J}.
$$
\gamma=\nu(2-\eta).
$$
Here we are however interested in the finite-size case with $L<\infty$.

\subsection{Critical case}
Let us assume that the system is critical at the yielding transition (as we assume to be the case for zero pinning concentration), which means that for $\gamma=\gamma_Y$ one has $\xi=L$ scaling-wise. Using Eq.~\eqref{eq:susc} it is then easy to prove that
\beq
\chi_{\rm peak} \propto L^{2-\eta},
\eeq
therefore the susceptibility peak for zero pinning concentration will grow like a power law of the size.

\subsection{Pinned case}
For every nonzero pinning concentration, we conversely assume the transition to be not critical, and that the lengthscale $\xi$ can grow at most up to the typical distance between pinned particles.
%At the susceptibility peak one would therefore have
%\beq
%\xi_{\rm peak}\simeq \frac{1}{c^{1/d}}.
%\label{eq:hyp}
%\eeq
In order to extract the dominant behavior for large $L$, we asymptotically expand the $\Gamma(x,a)$ for large $a\equiv L/\xi^*$, and then substitute Eq.~\eqref{eq:xistar} in the expansion. We obtain
\begin{widetext}
\beq
\chi_{\rm peak} \simeq f((\rho c)^{-1/d},\eta) - \exp\left(-(c\rho)^{1/d}L\right)((c\rho)^{-1/d})^{2-\eta}L^{1-\eta}\left[\left((\rho c)^{1/d}\right)^{1-\eta} - \frac{g((c\rho)^{-1/d},\eta)}{L} + O\left(\frac{1}{L^2}\right)\right],
\eeq
\end{widetext}
Where $f$ and $g$ are functions of $\xi^*=(\rho c)^{-1/d}$ and $\eta$ only. This implies that the peak height is given by a size-independent contribution, minus one with a negative exponential dependence on the size. In the pinned case we therefore expect that as the system size grows, the height of the susceptibility peak follows, at leading order in $\xi^*/L$,
\begin{equation}
 \chi_{\textrm{peak}} = A(c)-\exp[-(c\rho)^{1/d}L](c\rho)^{-1/d}L^{1-\eta} + O\left(\frac{\xi^*}{L}\right).
 \label{eq:theory}
\end{equation}
The susceptibility peak is therefore supposed to saturate from below to a finite value which depends only on the pinning concentration as $L\to\infty$, a manifestly non-critical behavior.
\begin{figure}
 \includegraphics[width=0.4\textwidth]{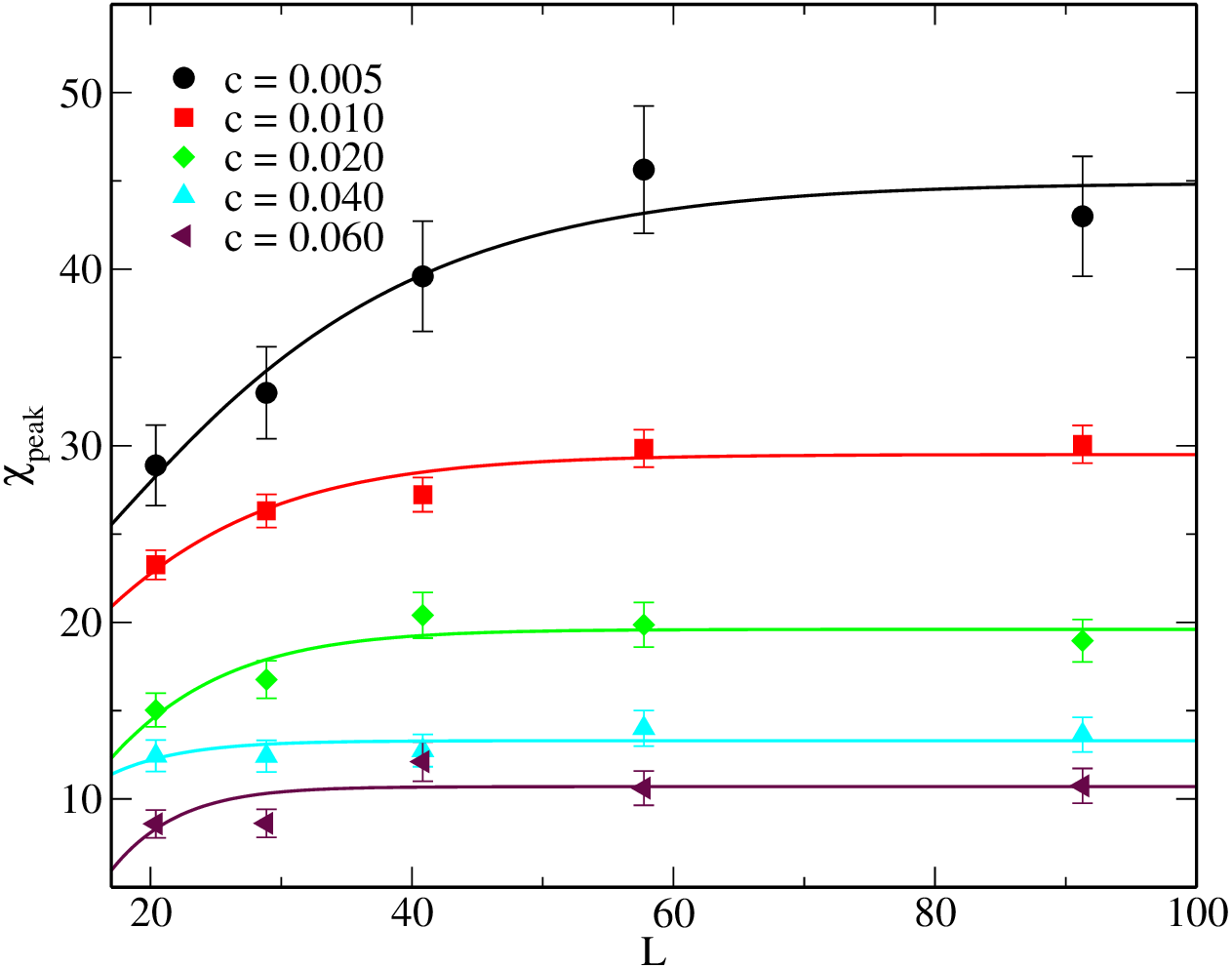}
 \caption{The susceptibility peaks from Fig.~\ref{fig:thesusc} fitted to eq.~\ref{eq:theory} as explained in the text. \label{fig:fits}}
\end{figure}
In Fig.~\ref{fig:fits} we test this hypothesis on the data reported in Fig.~\ref{fig:thesusc}, by fitting the peak height to the functional form~\eqref{eq:theory} (we fit A and $\eta$, and fix $c\rho$); the error bars are computed as follows: we first fit our theoretical formula, eq.~\eqref{eq:theory} to our simulation data; then for each parameter we have found an estimated percentage of error from the fitting program. The error bars are the computed by choosing
the parameter which had largest error, and multiplying it it by our simulation data for each $L$. The fits show satisfactory consistency between data and theory.

% the error bars amount to one standard deviation (i.e. patch-to-patch fluctuations) of the maximal susceptibility, and show how statistical convergence is more rapidly achieved at larger pinning concentrations. Nevertheless, our simple theory appears consistent with the data, also for $c=0.06$ when the error bars are the smallest.
% We remark that the best fit corresponds to $c=0.04$, and that a higher pinning concentration implies a more rapid statistical convergence of the correlators as apparent from Fig.~\ref{fig:thesusc}. We therefore ascribe the lower quality of the fits for lower concentrations to statistical noise.

\section{Discussion}
We have applied the critical picture (and associated tools)~\cite{16JPRS,17PPRS,17PRS} of the yielding transition to a system with inclusions in the form of pinned particles~\cite{CB12,KB13,OKIM15,CKD15-1,CKD15-2,DCK17}. As reported in~\cite{13DMPS,DCK17}, these inclusions have the effect of pinning and localizing the avalanches in the system (as well as raising the value of the yield strain $\gamma_Y$), transforming yielding into a crossover that takes place by progressive accumulation of localized plastic activity, as opposed to a transition that takes place when a scale-free avalanche sweeps the system inducing material failure~\cite{KLP10b}. We show that our formalism perfectly captures these features, and can explain them in terms of a very simple theory wherein the shear-banding correlation length is unable to grow beyond the typical distance between pinned particles. We remark that a similar way of suppressing shear-banding was proposed in ref.~\cite{HMPS16}, wherein the inclusions were instead modeled as elastic defects. A better understanding of the links between the two approaches would be a worthwhile research endeavor.

We also remark that a ``transition'' by progressive accumulation of localized avalanches is precisely the one found in the Random Field Ising Model~\cite{16NBT} when the disorder strength exceeds a critical threshold value~\cite{SDKKRS93}. This would suggest that, in terms of the picture proposed in~\cite{OBBRT18}, particle pinning has the result of effectively increasing the level of disorder in the system and thereby making it more ``ductile''. However, in other works~\cite{CB12,KB13}, pinning is used as a tool to increase the Kauzmann transition temperature $T_k$ of the system, which would on the contrary imply that pinned systems, by virtue of sitting closer to $T_k$, have a lower level of disorder compared to their unpinned counterparts, and are therefore supposed to have a more brittle behavior in terms of yielding and plasticity~\cite{OBBRT18}, which by our results and those of~\cite{PCK18}, is not the case. Resolving this apparent contradiction will be an obvious aim for future research, together with a study of behavior of the non-critical spinodal of the RFIM~\cite{16NBT} in presence of pinned spins, for the purpose of comparison with the results of this work.

\section{Acknowledgements}
This work had been supported in part by the Israel Science Foundation and the US-Israel Binational Science Foundation. SK would like to acknowledge the Weizmann Institute of Science Visiting Faculty program supported by ``Benoziyo Endowment Fund for the Advancement of Science''. Part of this work was carried out during SK's visit to Weizmann Institute of Science under the Visiting Faculty Program.

\appendix

\section{Expressions of the replica correlators\label{app:corrs}}
We report in this appendix the expression of the correlators we calculated in the main text. The details on how these expressions can be obtained can be found in~\cite{17PRS,17PPRS}. The correlation function is the sum of two terms:
\begin{equation}
 G_L(\boldsymbol{r}) \equiv 2G_R(\boldsymbol{r}) -\Gamma_2(\boldsymbol{r}),
 \label{eq:thecorr}
\end{equation}
with the first term being the so-called ``replicon'' correlator~\cite{Z10,98DKT}.
\begin{equation}
G_R({\bf r})  \equiv \frac{ \sum_{i\neq j}[u_i^{ab}u_j^{ab} - 2u_i^{ab}u_j^{ac} + Q_{ab} ~Q_{cd}]\delta({\bf r}-({\bf r}_{i}^a-{\bf r}_{j}^a)) }{ \sum_{i\neq j}\delta({\bf r}-({\bf r}_{i}^a-{\bf r}_{j}^a)) }.
 \end{equation}
and the second being, essentially, a four-point correlation function:
\begin{equation}
\Gamma_2(\boldsymbol{r})\equiv \frac{ \sum_{i\neq j}(u^{ab}_i-Q_{ab}) (u^{ab}_j-Q_{ab})\delta(\boldsymbol{r}-(\boldsymbol{r}_{i}^a-\boldsymbol{r}_{j}^a)) }{ \sum_{i\neq j}\delta(\boldsymbol{r}-(\boldsymbol{r}_{i}^a-\boldsymbol{r}_{j}^a)) },
\end{equation}
with the definition
\begin{equation}
  u^{ab}_i \equiv \theta(\ell-|{\bf r}_i^a-{\bf r}_i^b|) \ ,
 \end{equation}
 and $Q_{ab}$ is the global order parameter first defined in~\cite{16JPRS}
 \begin{equation}
 Q_{ab} \equiv \frac{1}{N}\sum_{i=1}^N \theta\left(\ell-\left| \boldsymbol{r}_i^a - \boldsymbol{r}^b_i\right| \right),
 \end{equation}
 and $\theta(x)$ is the Heaviside step function. We take $\ell=0.35$ in LJ units, following~\cite{16JPRS,17PRS,17PPRS}.\\ 
These quantities are computed for each unique quadruplet of replicas in a single patch, for each value of the strain $\gamma$, and the result is afterwards averaged over the patches as detailed in the main text. The associated susceptibility, which we report in this work, is simply the integral over space of eq.~\eqref{eq:thecorr}.

\bibliographystyle{mioaps}
\bibliography{LJ,bibliografia}

\end{document}